\documentclass[12pt,preprint]{aastex}

\slugcomment{To appear in $The$ $Astrophysical$ $Journal$}

\shorttitle{Investigation for the puzzling abundance pattern in CS
30322-023} \shortauthors{Cui et al.}

\begin{document}
\title {Investigation for the puzzling abundance pattern of the neutron-capture elements in the ultra metal-poor star: CS 30322-023}
\author{Wen-Yuan Cui\altaffilmark{1,2}, Bo Zhang\altaffilmark{1,2,4}, Kun Ma\altaffilmark{1} and Lu Zhang\altaffilmark{3}}

\affil{$^{1}$Department of Physics, Hebei Normal University, 113
Yuhua Dong Road, Shijiazhuang 050016, P.R.China;}
\email{zhangbo@hebtu.edu.cn}

\affil{$^{2}$National Astronomical observatories, Chinese Academy
of Sciences, 20A Datun Road, Chaoyang District, Beijing 100012,
P.R.China;}\email{wenyuancui@126.com}

\affil{$^{3}$Department of Modern Physics, University of Science
and Technology of China, Hefei 230026, P.R.China;}
\email{zhang8899$_{-}$2000@163.com}

\altaffiltext{4}{ Corresponding author. E-mail address:
zhangbo@hebtu.edu.cn}

\begin{abstract}
The s-enhanced and very metal-poor star CS 30322-023 shows a
puzzling abundance pattern of the neutron-capture elements, i.e.
several neutron-capture elements such as Ba, Pb etc. show
enhancement, but other neutron-capture elements such as Sr, Eu
etc. exhibit deficient with respect to iron. The study to this
sample star could make people gain a better understanding of s-
and r-process nucleosynthesis at low metallicity. Using a
parametric model, we find that the abundance pattern of the
neutron-capture elements could be best explained by a star that
was polluted by an AGB star and the CS 30322-023 binary system
formed in a molecular cloud which had never been polluted by
r-process material. The lack of r-process material also indicates
that the AGB companion cannot have undergone a type-1.5 supernova,
and thus must have had an initial mass below $4.0$M$_\odot$, while
the strong N overabundance and the absence of a strong C
overabundance indicate that the companion's initial mass was
larger than $2.0$M$_\odot$. The smaller s-process component
coefficient of this star illustrates that there is less accreted
material of this star from the AGB companion, and the sample star
should be formed in the binary system with larger initial orbital
separation where the accretion-induced collapse (AIC) mechanism
can not work.\end{abstract}


\section{Introduction}
The two neutron-capture process, the slow (s-process), and the
rapid (r-process), occur under different physical conditions, and
therefore are likely to arise in different astrophysical sites.
The dominant site of the s-process is thought to be the Asymptotic
Giant Branch (AGB) phase in low- and intermediate-mass stars
\citep{bus99}. The site or sites of the r-process are not known,
although suggestions include the $\nu$-driven wind of Type II
supernova \citep{woo92,woo94}, the mergers of neutron stars
\citep{lat74,ros00}, AIC \citep[accretion-induced
collapse;][]{qia03}, and Type-1.5 supernova \citep{ibe83,zij04}.
The neutron-capture elements can be composed of some pure
r-process, pure s-process, and some mixed-parentage isotopes. As a
result, when the solar system total abundances(t$_{ss}$) are
separated into contribution from the s-process(s$_{ss}$) and the
r-process(r$_{ss}$), some elements are mostly contributed by
r-process, such as Eu, and some by the s-process, such as Ba.
Therefore Eu is commonly referred to as an "r-process element" and
Ba, as an "s-process element".

In an early systematic study of elemental abundances in halo
stars, \citet{spit78} found that the neutron-capture elements Ba
and Y were over-deficient with respect to iron, and that the
barium abundance increased almost as fast as iron at low
metallicity, suggesting a common origin in massive stars. Based on
this suggestion, \citet{tru81} supposed that in the early Galaxy
the neutron-capture elements were formed exclusively through the
r-process. Observations of metal-poor stars with metallicities
lower than [Fe/H] $\approx-2.5$ enriched in neutron-capture
elements have further supported this hypothesis
\citep{sne96,sne03,bur00,cay01,hil02}. Although the material from
which \ion{Pop}{2} stars form is not expected to contain
significant s-process contributions, some stars including some
subgiants are greatly enriched in carbon and s-elements
\citep[s-rich stars hereafter;][]{nor97,hil00}. These are believed
to be binary companions of initially more-massive donor stars
which have evolved through the thermally pulsing AGB phase and
transferred material enriched in C and s-process elements onto the
lower mass, longer lived secondary now observed.

The generally favoured s-process model by now is associated with
the partial mixing of protons (PMP hereafter) into the radiative
C-rich layers during thermal pulses
\citep{gal98,gal03,str95,str05}. PMP activates the chain of
reactions $^{12}$C(p,$\gamma$)$^{13}$N($\beta$)\\
$^{13}$C($\alpha$,n)$^{16}$O which likely occurs in a narrow mass
region of the He intershell (i.e. $^{13}$C-pocket) during the
interpulse phases of an AGB star. Using the primary-like neutron
source (i.e. $^{13}$C($\alpha$,n)$^{16}$O) and starting with a
very low initial metallicity, most iron seeds are converted into
$^{208}$Pb. So, when third dredge-up episodes mix the
neutron-capture products into the envelope, the star appears
s-enhanced and lead-rich. The nucleosynthesis of neutron-capture
elements in the Carbon-Enhanced Metal-Poor stars \citep[hereafter
CEMP stars;][]{coh05} can be investigated by abundance studies of
s-rich or r-rich stars. Recently, \citet{mas06} analyzed the
spectra of the lead-rich ultra-metal-poor star CS 30322-023, and
concluded that the observed abundances could not be well fit by a
scaled solar system r-process (r$_{ss}$) pattern nor by the
s-process pattern of an AGB model (0.8M$_{\odot}$, [Fe/H]
$=-3.8$). This star shows enhancement of the neutron-capture
elements Ba and Pb, but the discovery that several neutron-capture
elements such as Sr and Eu are deficient with respect to iron is
puzzling. \citet{mas06} estimated that CS 30322-023 is the most
evolved CEMP star, probably on the AGB stage, possibly even in the
thermally pulsing (helium shell flashing) stage.  However, there
is another possibility for the s-enhanced scenario of CS
30322-023. Because of the strong N overabundance, the Na
overabundance and the absence of a strong C overabundance,
\citet{mas06} have also speculated that this star should be
polluted by an AGB star, which has a mass of at least
2.0M$_{\odot}$ for the possible Hot-bottom burning. Clearly, the
restudy of elemental abundances in this object is still very
important for well understanding the nucleosynthesis of
neutron-capture elements in early Galaxy.

There have been many theoretical studies of s-process
nucleosynthesis in low-mass AGB stars. Unfortunately, the precise
mechanism for chemical mixing of protons from the hydrogen-rich
envelope into the $^{12}$C-rich layer to form a $^{13}$C-pocket is
still unknown \citep{bus01}. It is interesting to adopt the
parametric model for metal-poor stars presented by \citet{aoki01}
and developed by \citet{zha06} to study the physical conditions
which could reproduce the observed abundance pattern found in this
star. In this paper we investigate the characteristics of the
nucleosynthesis pathway that produces the special abundance ratios
of s-rich object CS 30322-023 using the s-process parametric model
\citep{zha06}. The calculated results are presented in Sect.\ 2 in
which we also discuss the characteristics of the s-process
nucleosynthesis at low metallicity. Conclusions are given in
Sect.\ 3.

\section{Results and Discussion}
We explored the origin of the neutron-capture elements in CS
30322-023 by comparing the observed abundances with predicted s-
and r-process contribution. For this purpose, we adopt the
parametric model for metal-poor stars presented by Zhang et
al.(2006). The $i$-th element abundance in the envelope of the
star can be calculated as follows:
\begin{equation}
N_{i}(Z)=C_{s}N_{i,\ s}+C_rN_{i,\ r}10^{[Fe/H]} ,
\end{equation}
where $Z$ is the metallicity of the star, $N_{i,\ s}$ is the
abundance of the \textit{i}-th element produced by the s-process
in the AGB star and $N_{i,\ r}$ is the abundance of the
\textit{i}-th element produced by the r-process (per Si $=10^6$ at
$Z=Z_\odot$), $C_s$ and $C_r$ are the component coefficients that
correspond to contributions from the s-process and the r-process
respectively.

There are four parameters in the parametric model of s- and r-rich
stars (hereafter s+r stars). They are the neutron exposure per
thermal pulse $\Delta\tau$, the overlap factor $r$, the component
coefficient of the s-process $C_{s}$ and the component coefficient
of the r-process $C_{r}$. The adopted initial abundances of seed
nuclei lighter than the iron peak elements were taken to be the
solar-system abundances, scaled to the value of [Fe/H] of the
star. Because the neutron-capture-element component of the
interstellar gas that formed very mental-deficient stars is
expected to consist of mostly pure r-process elements, for the
other heavier nuclei we use the r-process abundances of the solar
system \citep{arla99}, normalized to the value of [Fe/H]. The
abundances of r-process nuclei in equation (1) are taken to be the
solar-system r-process abundances \citep{arla99} for the elements
heavier than Ba, for the other lighter nuclei we use solar-system
r-process abundances multiplied by a factor of 0.4266
\citep{zha06}. Using the observed data in the sample star CS
30322-023 \citep{mas06}, the parameters in the model can be
obtained from the parametric approach.

Figure 1 shows our calculated best-fit result. For this star, the
curves produced by the model are consistent with the observed
abundances within the error limits. The agreement of the model
results with the observations provides strong support to the
validity of the parametric model. In the AGB model, the overlap
factor, $r$, is a fundamental parameter. \citet{gal98} have found
an overlap factor of $r \simeq 0.4 - 0.7$ in their standard
evolution model of low-mass ($1.5-3.0M_{\odot}$) AGB stars at
solar metallicity. The overlap factor calculated for other
s-enhanced metal-poor stars lies between 0.1 and 0.81
\citep{zha06}. The overlap factor deduced for CS 30322-023 is
about $r=0.65^{+0.10}_{-0.19}$, which lies in both ranges above.

The neutron exposure per pulse, $\Delta\tau$, is another
fundamental parameter in the AGB model. \citet{zha06} have deduced
the neutron exposure per pulse for other s-enhanced metal-poor
stars which lies between 0.45 and 0.88 mbarn$^{-1}$. The neutron
exposure deduced for CS 30322-023 is about $\Delta\tau =
0.55^{+0.18}_{-0.06}$ mbarn$^{-1}$. In the case of multiple
subsequent exposures, the mean neutron exposure is given by
$\tau_{0}= -\Delta\tau/ln\textit{r}$. It is noteworthy, however,
that the value of
$\tau_{0}=(2.38^{+0.91}_{-0.57})(\frac{T_{9}}{0.348})^{1/2}$
mbarn$^{-1}$(in this work, $T_{9}=0.1$, in units of 10$^{9}$ K)
for CS 30322-023 is significantly larger than those which best fit
solar-system material,
$\tau_{0}=(0.30\pm0.01)(\frac{T_{9}}{0.348})^{1/2}$ mbarn$^{-1}$
\citep{kap89}. In fact, the higher mean neutron exposure favors
large amounts of production of the heavier elements such as Pb, Ba
etc. and less Sr, Y etc. \citep{cui06}, which should be one reason
of the puzzling abundance pattern of the s-process elements, i.e.
Ba, Pb etc. show enhancement, and Sr, Y etc. exhibit deficient
with respect to iron.

We explore possibility that CS 30322-023 belongs to a binary
system. In this case, the enhancement of the neutron-capture
elements Ba and Pb suggests that in a binary system a
mass-transfer episode from a former AGB star took place. One major
goal of this work is to explore the astrophysical condition,
associated with an AGB star in a binary system, in which several
neutron-capture elements such as Sr and Eu are deficient with
respect to iron. It should be noted that the s-process abundances
in the envelope of the stars could be expected to be lower than
the abundance produced by the s-process in the AGB star because
the material is mixed with the envelopes of the primary (former
AGB star) and secondary stars. The component coefficient of the
r-process calculated for CS 30322-023 is about $C_{r}\sim0$, which
means that the star and its AGB companion formed in a molecular
cloud which had never been polluted by r-process material. This
fact can directly lead to the special abundance pattern that
several neutron-capture elements are deficient with respect to
iron, such as Eu and Sr. \citet{zij04} speculates that the strong
metallicity dependence of mass loss during the AGB phase leads to
a steeper initial-final mass relation for low-metalliciy stars;
that is, for a given initial mass the final mass is higher for
metal-poor stars. Therefore, the core of, e.g., an metal-poor star
with [Fe/H]$=-3.0$ having an initial mass of 4.0M$_{\odot}$, would
reach the Chandrasekhar mass, leading to a "Type 1.5 supernova".
In such a supernova, r-process nucleosynthesis might have
occurred, and the surface of the companion star observed today
could have been polluted with the elements that were produced.
Since the sample star had never been polluted by r-process
material, a smaller initial mass (M$<4.0$M$_{\odot}$) of the
former AGB companion could be expected. In addition, based on the
strong N overabundance, the Na overabundance and the absence of a
strong C overabundance, \citet{mas06} have speculated that CS
30322-023 should be polluted by an AGB star, which has a mass of
at least 2.0M$_\odot$.

\citet{qia03} proposed a theory for the creation of s+r stars.
They suggest that a s+r star is a member of a binary system in
which the former primary went through the AGB phase. In this
stage, carbon and s-process elements were dumped onto the surface
of the companion via mass transfer. The former primary then
evolved into a white dwarf. Mass transfer then occurred in the
reverse direction, i.e., from the companion to the white dwarf,
leading in turn to an accretion-induced collapse (AIC) of the
white dwarf into a neutron star. Subsequent r-process
nucleosynthesis is then presumed to occur in a neutrino-driven
wind of the neutron star, and the nucleosynthesis products
contaminate the surface layers of the star that is observed today.

The binary system with lower-mass AGB stars (M$<$4.0M$_{\odot}$)
and larger initial orbital separation could not cause AIC, because
the white dwarf accretes matter insufficiently from the polluted
star and the star is polluted only by the AGB companion, which can
explain the formation of s-only stars. The r-process component
coefficient of CS 30322-023 is about 0, which implies that this
star belongs to s-only star. \citet{zha06} have calculated two
other s-only stars with $C_{s}=0.0005$ and 0.0006 and 10 s+r stars
with $0.0017\leq C_{s}\leq0.0060$. The s-process component
coefficient of CS 30322-023 is about 0.0002 which is smaller than
the other s-rich stars. This fact illustrates that there is less
accreted material of CS 30322-023 from the AGB companion, and the
sample star should be formed in the binary system with a larger
initial orbital separation which could not cause AIC. This should
be another reason which leads to the special abundance pattern
that several neutron-capture elements are deficient with respect
to iron, such as Sr and Y. In addition, the s-process pattern of
this star is matched by a model with $r=0.65$, which is large
enough to be consistent with the idea that the AGB star polluting
the sample star was of low mass
\citep[M$<$4.0M$_{\odot}$;][]{cui06}.

As a star evolves off the main sequence, its convective envelope
penetrates deeper and deeper, reaching into regions where there
was previous H burning, and thus undergoes the so-called first
dredge-up. Later on, when a star evolves off the red giant branch
(RGB), a further mixing episode can take place (usually referred
to as "extra mixing"). \citet{mas06} have estimated that CS
30322-023 is an evolved CEMP star. As such, it would be expected
to have passed through the deepest convective envelope stage,
i.e., the amount of accreted material from the former AGB star,
including C and s-elements such as Sr and Ba et al., is further
diluted by the material of the stellar convective envelope, thus
decreasing both [C/Fe] and [s/Fe]("s" representing a specie of
s-process element). Although the mixing processes (first dredge-up
and extra mixing) should have occurred in the sample star, the
effect of orbital separation of the binary system above mentioned
is expected to be also important.

It is interesting to estimate the dilution effect brought by the
orbital separation quantitatively for CS 30322-023 with
$C_{s}=0.0002$. The s-process abundances of the sample star are a
result of pollution from the dredged-up material of the former AGB
star. The measured [s/Fe] refers to the average s-processed
material of the AGB star after dilution by mixing with the
envelope of the little companion that is now the extrinsic star.
At some time on the AGB, the convective He-shell and the envelope
of the giant will be overabundant in heavy element respectively by
a factor $f_{shell}$ and $f_{env,\ 1}$ with respect to solar
system abundances normalized to the value of [Fe/H]. The
approximate relation between $f_{env,\ 1}$ and $f_{shell}$ is
\begin{equation}
f_{env,\ 1}\approx\frac{\Delta M_{dr}}{M^e_1}f_{shell} ,
\end{equation}
where $\Delta$$M_{dr}$ is the total mass dredged up from the
He-shell into the envelope of the AGB star and $M^{e}_{1}$ is the
envelope mass of the AGB star. For a given s-process element, the
overabundance factor $f_{env,\ 2}$ in the future s-enhanced star
envelope can be approximately related to the overabundance factor
$f_{env,\ 1}$ by
\begin{equation}
 f_{env,\ 2}\approx\frac{\Delta M_{2}}{M^e_2}f_{env,\ 1}
 \approx\frac{\Delta M_2}{M^e_2}\frac{\Delta M_{dr}}{M^e_1}f_{shell} ,
\end{equation}
where $\Delta$$M_{2}$ is the amount of matter accreted by the
future s-enhanced star, $M^{e}_{2}$ is the envelope mass of the
star. The component coefficient $C_{s}$ is computed from the
relation
\begin{equation}
 C_s =\frac{f_{env,\ 2}}{f_{shell}}\approx\frac{\Delta M_2}{M^e_2}\frac{\Delta
 M_{dr}}{M^e_1} .
\end{equation}

For wide binary systems mass transfer operates through stellar
winds, rather than by Roche-Lobe overflow \citep{bof88}. We use
the Bondi-Hoyle accretion rate to calculate the mass being
accreted by the sample star, \citep{the96,jor98}:
\begin{equation}
 \Delta M_2=-\frac{\alpha}{A^2}\Bigg[\frac{GM_2}{\upsilon_{ej}^2}\Bigg]^2
 \Bigg[\frac{1}{1+(\upsilon_{orb}/\upsilon_{ej})^2}\Bigg]^{\frac{3}{2}}\Delta M_1 ,
\end{equation}
where $A$ is the semi-major axis of the orbit of the binary
system, $\alpha$ is a constant expressing the accretion
efficiency,  and  $\alpha=0.1$ in the situation of interest here
according to the detailed hydrodynamic simulations
\citep{the96,jor98}. $\upsilon_{ej}$ is the wind velocity and
$\upsilon_{orb}$ is the orbital mean velocity. $\Delta M_1$ is the
wind mass-loss of the former AGB star in a binary system, i.e.,
the present white dwarf star.

For the evolved sample star \citep[$\sim0.8M_\odot$;][]{mas06},
the mass of the convective envelope, $M_2^{e}$ ,is larger than the
main-sequence stars with similar stellar mass. As a result, the
dilution of the accreted material is relatively large and we take
$M_2^{e}\approx 0.5$M$_\odot$ \citep{bof88}. Adopting the formulas
(4) and (5) and taking $\Delta$$M_{dr}\approx 0.03-0.05$M$_\odot$
\citep{gal98}, $\Delta M_1\approx M^e_1$, we can estimate the
fraction of the mass captured by the little companion ($\sim
$0.8M$_\odot$) to be about from $0.2\%$ to $0.3\%$ of the mass
lost by the wind (15 km s$^{-1}$) from the former AGB star ($\sim
$3.0M$_\odot$) with a period of about $10-20$ yr. This implies
that the sample star, CS 30322-023, should belong to a long-period
system, consistent with what is expected that the accretion rate
is smaller in wider systems. Because of the uncertainties of
mass-loss rates and our poor knowledge of how and when mass
transfer phenomena occur, we do not claim that this is the only or
even the best understanding of the s-process component
coefficient. Obviously, long-term radial-velocity monitoring is
advisable, possibly leading to the determination of orbital
elements for CS 30322-023, which could allow for a further test of
our results.

It was also possible to isolate the contributions corresponding to
the s- and r-process by our parametric model. In the Sun, the
elemental abundances of Ba and Eu consist of significantly
different combinations of s- and r-process isotope contributions,
with s:r ratios for Ba and Eu of 81:19 and 6:94, respectively
\citep{arla99}. The Ba and Eu abundances are most useful for
unraveling the sites and nuclear parameters associated with the s-
and r-process correspond to those in extremely metal-poor stars,
polluted by material with a few times of nucleosynthetic
processing. We explored the contributions of s- and r-process for
these two elements in the sample star. From equation (1), we
obtained the s:r ratios for Ba and Eu are 99.6:0.4 and 74.2:25.8,
which are obviously larger than the ratios in the solar system.
The abundances of r-elements, such as Eu, in the sample star is
lower than those in normal stars, so the sample star seemed to be
an "extremely" s-only star.

We discuss the uncertainty of the parameters using the method
presented by \citet{aoki01}. We choose Sr as the representative
for the first peak elements, Ba as the representative for the
second peak elements, Pb as the representative for the third peak
elements and Eu as the representative for the r-process elements.
There are two regions in Figure 2, where the region II is
$\Delta\tau = 0.55^{+0.18}_{-0.06}$ mbarn$^{-1}$ and
$r=0.65^{+0.10}_{-0.19}$ with $\chi^2=0.556$, the region I is
$\Delta\tau = 0.19^{+0.09}_{-0.12}$ mbarn$^{-1}$ and
$r=0.87^{+0.10}_{-0.03}$  with $\chi^2=0.846$, in which all the
observed ratios of four representative elements can be accounted
for within the error limits. The corresponding mean neutron
exposure for the region I and II are $\tau_0 =
1.36^{+0.68}_{-0.23}$ mbarn$^{-1}$ and $\tau_0 =
1.28^{+0.91}_{-0.57}$ mbarn$^{-1}$, respectively. Since the final
abundance distributions of the s-process nucleosynthesis depend
mainly upon the mean neutron exposure $\tau_0$, we investigate the
relation between the mean neutron exposure $\tau_0$ and the
overlap factor $r$. It is noteworthy that although the values of
the overlap factor and the neutron exposure are obviously
different for the two regions, indeed, the value of the mean
neutron exposure for region I is close to that of region II. The
uncertainties of the parameters for the star CS 30322-023 are
similar to those for other metal-poor stars obtained by
\citet{zha06}.

\section{ Conclusions}
The star CS 30322-023 is an ultra metal-poor lead star which has a
special abundance pattern of the neutron-capture elements, i.e.
this star shows enhancement of the neutron-capture elements Ba and
Pb, but several neutron-capture elements such as Sr and Eu are
deficient with respect to iron. The neutron-capture elements
abundance pattern could be explained by a binary system which
should form in a molecular cloud which had never been polluted by
r-process material. This fact can explain that several
neutron-capture elements are deficient with respect to iron, such
as Eu and Sr partly. The initial mass of the AGB companion should
be smaller than $4.0$M$_\odot$, which excludes the possibility of
forming a type-1.5 supernova, and larger than $2.0$M$_\odot$
because of the strong N overabundance and the absence of a strong
C overabundance. The r-process component coefficient of CS
30322-023 is very small, which implies that this star is an
"extreme" s-only stars. The s-process component coefficient of
this star is smaller than the other s-rich stars. Although the
mixing processes (first dredge-up and extra mixing) should have
occurred in the sample star, the effect of orbital separation of
the binary system is also important. The sample star CS 30322-023
should belong to a wide binary system where the AIC mechanism can
not work. This should be an important reason which leads to the
special abundance pattern that several neutron-capture elements
are deficient with respect to iron, such as Sr, Y and Eu.
Certainly, the possibility that CS 30322-023 should be a TP-AGB
star \citep{mas06} is also existent. In this case, the operation
of the s-process is associated with thermal pulses occurring on
the AGB. As a result of the so-called "third dredge-up", s-process
enriched material is brought to the surface of the AGB star.
Obviously, large uncertainties still remain in this topic, and
full understanding of the abundance pattern and evolutionary state
of CS 30322-023 will depend on future high-resolution studies and
long-term radial-velocity monitoring. More in-depth theoretical
and observational studies of s-rich stars will reveal the
characteristics of the s-process at low metallicity and the
history of enrichment of s- and r-elements in the early Galaxy.

\acknowledgments

We thank the referee for an extensive and helpful review,
containing very relevant scientific advice which improved this
paper greatly. This work has been supported by the National
Natural Science Foundation of China under grant no.10673002.

\clearpage


\begin{figure}
\epsscale{.80} \plotone{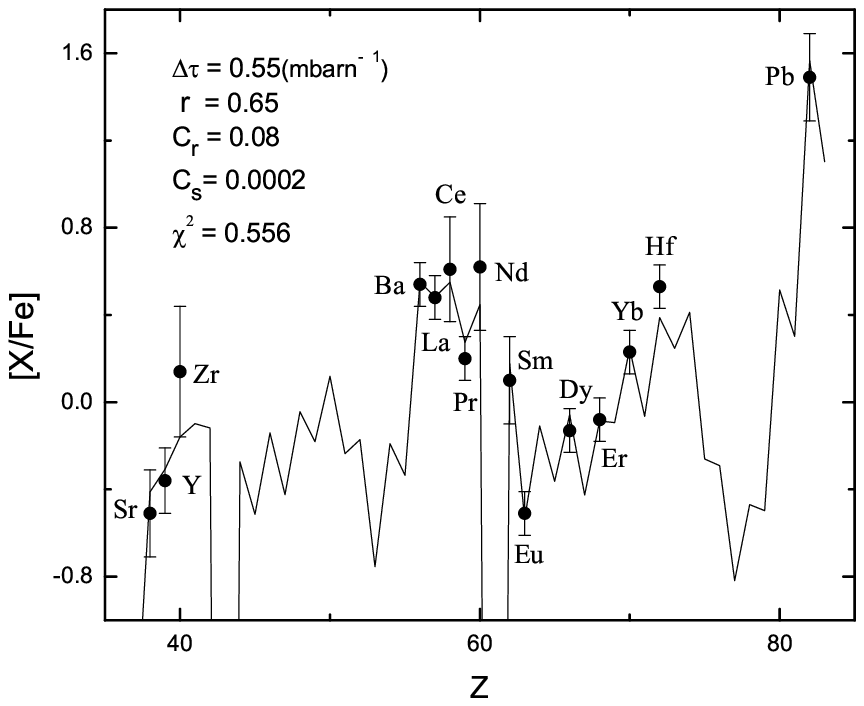} \caption{{Best fit to
observational result of metal-deficient star CS 30322-023. The
black circles with appropriate error bars denote the observed
element abundances, the solid lines represent predictions from
s-process calculations considering r-process contribution. The
standard unit of $\Delta\tau$ is mbarn$^{-1}$. }\label{fig1}}
\end{figure}

\begin{figure}
\epsscale{.80} \plotone{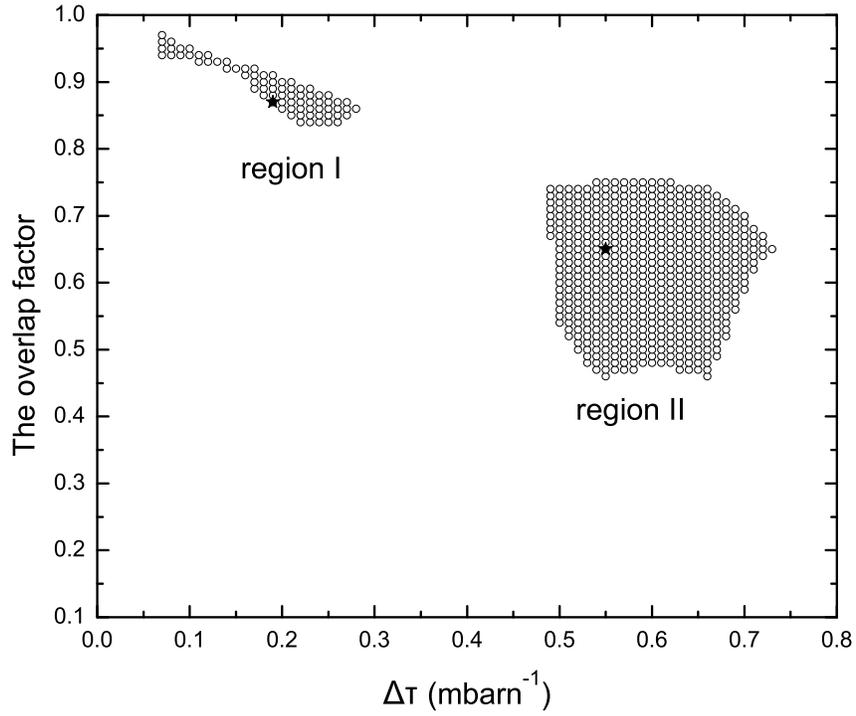} \caption{{The possible
combinations (unfilled circles) of neutron exposure and overlap
factor by which all the observed ratios of four representative
elements, i.e. Sr, Ba, Eu, Pb, can be accounted for within the
error limits. The two filled asterisks correspond respectively the
combinations $\Delta\tau = 0.55$ mbarn$^{-1}$, $r=0.65$ with
$\chi^2=0.556$ (the minimum value of $\chi^2$), and $\Delta\tau =
0.19$ mbarn$^{-1}$, $r=0.87$ with $\chi^2=0.846$ (the minimum
value of $\chi^2$ in the region I ). }\label{fig2}}
\end{figure}


\end{document}